\documentclass[a4paper]{article}
\usepackage[english]{babel}
\usepackage[utf8x]{inputenc}
\usepackage[T1]{fontenc}

\usepackage[a4paper,top=3cm,bottom=2cm,left=3cm,right=3cm,marginparwidth=1.75cm]{geometry}
\usepackage{amsmath}
\usepackage{graphicx}

\DeclareMathOperator*{\argmin}{arg\,min}

\title{Real-time single-pixel video imaging with Fourier domain regularization} 
\author{Krzysztof M. Czajkowski, Anna Pastuszczak, \\ Rafa{\l} Koty{\'n}ski}
\date{University of Warsaw, Faculty of Physics, Pasteura 5, 02-093 Warsaw, Poland}

\begin{document}

\maketitle




\begin{abstract}
We present a closed-form image reconstruction method for single pixel imaging based on the generalized inverse of the measurement matrix. Its numerical cost scales linearly with the number of measured samples. Regularization is obtained by minimizing the norms of the convolution between the reconstructed image and a set of spatial filters, and the final reconstruction formula can be expressed in terms of matrix pseudoinverse.
At high compression this approach is an interesting alternative to the methods of compressive sensing based on l1-norm optimization, which are too slow for real-time applications.
For instance, we demonstrate experimental single-pixel detection with real-time reconstruction obtained in parallel with the measurement at the frame rate of $11$ Hz for highly compressive measurements with the resolution of $256\times 256$. For this purpose, we 
preselect the sampling functions to match the average spectrum obtained with an image database. The sampling functions are selected from the Walsh-Hadamard basis, from the discrete cosine basis, or from a subset of Morlet wavelets convolved with white noise. We show that by incorporating the quadratic criterion into the closed-form reconstruction formula, we are able to use binary rather than continuous sampling reaching similar reconstruction quality as is obtained by minimizing the total variation. This makes it possible to use cosine or Morlet-based sampling with digital micromirror devices without advanced binarization methods. 
\end{abstract}


\section{Introduction}
Indirect compressive imaging techniques termed as single pixel imaging and computational ghost imaging~\cite{Baraniuk2008, PRA_78_061802_Shapiro}
contribute to many novel ideas in optics. Prospect applications of these measurement methods include spectral imaging~\cite{Bian:scirep_6_24752,Li:scirep_7_41435,Multispectral_scirep2017Jin,Multispectral_life_nphot2017Pian}, polarimetric imaging~\cite{OL_37_824_Duran,Soldevila2013,Polarimetric_AO2018Fade}, 3D imaging~\cite{Science_Sun2013,Sun:natcommun_7_12010,Li:ao_53_7992}, around-the-corner imaging~\cite{Zhang:natcom_6_6225}, imaging through scattering media~\cite{Duran:15}, imaging at low light-levels~\cite{Liu_SciRep_8_5012_2018}, spectroscopy~\cite{Spectroscopy_AO2016Starling}, pattern recognition~\cite{pastor:oc2017} etc. The inherently indirect principle of the measurement in single pixel detectors  makes it necessary to reconstruct the image numerically after the measurement. This is computationally demanding, and for compressive measurements, the reconstruction is  ambiguous and typically involves optimization methods. The framework of compressive sensing~\cite{IEEE_SPM_5_21_Candes,IEEE_SPM_25_14_Romberg,fast_l1_ieee2018Donoho,Sampling_theory_Eldar,Bian:josaa_35_78_2018}
offers methods that give a very good reconstruction quality and are based on constrained $\ell^1$-norm minimization of a sparse image representation. On the other hand, single step reconstruction methods with regularization~\cite{Bian:josaa_35_78_2018,Dokmanic_ieee_icassp_6638923_2013} used to solve the ill-posed inverse problem are a lot faster, at least for moderately sized measurements. 

One of the greatest challenges for the single-pixel cameras is real-time imaging at a video rate \cite{video_vis_ir_scirep2015Edgar,video_gas_OE2017Gibson,Ultrasound_optica2016Huynh,Fast_Fourier_scirep2017Zhang,Deep_learning_scirep2018Higham}.  
The video frame rate is limited by the method of sampling itself, requiring a spatial light modulator (SLM) to be switched to different states as many times as the number of captured image samples. For a compressive measurement this number is usually a small fraction of the pixel resolution of the image. The state-of-the-art SLMs offer sampling rates above $20$~kHz. The modulation speed of structured illumination in computational ghost imaging set-ups may be further increased by combining various light modulation techniques together or by using arrayed light sources with a fast modulation rate~\cite{Galvanic_scirep2017Wang,Xu_oe_26_2427_2018,Liu_oe_26_10048_2018}.
However, iterative optimization algorithms are usually not fast enough to allow for real-time image reconstruction at the similar rate. Therefore, efficient single-step reconstruction methods for single-pixel video imaging, providing reasonable quality of the recovered images, are needed.

 In this paper we focus on the closed-form solutions to the inverse problem, which may be cast to the calculation of the Moore-Penrose pseudoinverse. 
 The product of the pseudoinverse of the measurement matrix and the vector containing measurements captured by a single-pixel camera already provides an estimate for the measured image, and has been used for single pixel detection in conjunction with various nonorthogonal sampling protocols~\cite{OE_22_30063_Zhang2014,Gong_photonres_3_234_2015,Czajkowski:scirep_8_466}. 
 On the other hand, when the measurement patterns represent a subset of functions of an orthogonal or unitary transform, the pseudoinverse based solution is equivalent to that obtained with a respective inverse transform from the incomplete data. While for an overdetermined linear system, the pseudoinverse may be used to find the minimal mean-square-error solution, in the case of an underdetermined system it gives the solution with the minimal Euclidean norm. This kind of regularization is not necessarily what one needs in single-pixel detection.
We refine this result by including minimization of quadratic criterion into the solution. The criterion considered by us combines the minimization of the image gradient with a penalty-function for high spatial frequency components of the image. It is expressed as a $\ell^2$ norm of the convolution between the image and a symmetrical filter proposed in this work. The final result is also a closed-form pseudoinverse-based method. It shares the advantage of the direct pseudoinverse based reconstruction in terms of its linear numerical cost as a function of the dimension of the measurement.
It makes sense to precalculate the reconstruction matrix  before the measurements takes place and to store it in memory. This approach is feasible for a reasonably small dimension of the measurement vector, typically on the order of several thousands elements. For the resolution of $256\times 256$, $2000$ elements correspond to the compression ratio of around $3\%$. 
  With a $22$~kHz digital micromirror device (DMD), this allows for compressive single-pixel detection at the rate of over $11$~Hz with image reconstruction on the fly.

\section{Image reconstruction algorithm}
Single-step image reconstruction and a linear cost of the reconstruction algorithm as a function of the size of the compressive measurement are two factors desired for real time single pixel imaging. 
In our case, the reconstructed image is obtained as the solution to a minimization of a quadratic criterion with a linear constraint. The mathematical formulation of the reconstruction algorithm is presented below. Further in the text we will call it the Fourier domain regularized inversion method (FDRI).  
The usual single pixel detection measurement model is assumed,
\begin{equation}
M\cdot x=y,\label{eq.measurement}
\end{equation}
where $x$ is the measured image consisting of  $n$  real-valued pixels, $y$ is the result of measurement that includes $k$ elements, and $M$ is a rectangular $k\times n$ measurement matrix.  
  Rows of $M$ include the sampling patterns, and  $k<n$ for a compressive measurement. 
The image is reconstructed from the measurement $y$ through minimization of a quadratic criterion constrained by Eq.~(\ref{eq.measurement}),
\begin{equation}
 x_0 =\argmin_{x} E(x)\textnormal{ subject to }M\cdot x=y.\label{eq.optim}
\end{equation}
The proposed criterion is
defined as the squared $\ell^2$ norm of the (circular) convolution of $x$ and a filter $h$,
\begin{equation}
E(x)= \| x * h \|^2.\label{eq.crit}
\end{equation}
The same criterion (\ref{eq.crit}) may be expressed as a quadratic form,
\begin{equation}
E(x)=x^{T}\cdot C\cdot x={\hat x}^{T}\cdot \hat{C}\cdot {\hat x},
\end{equation}
where $C$ is a circulant positive semi-definite square matrix. Its Fourier transform $\hat C=F\cdot C\cdot F^{*}$ is diagonal and consists of  non-negative elements ${\hat C}_{i,j}=\delta_{i,j} |\hat h_i|^2$, where $\hat h$ is the transfer function of the filter $h$. We use the caret to denote the Fourier transformed vectors and matrices, and $F$ is the respective 2D Fourier transform matrix. Moreover, we
assume that  $|\hat h_i|^2$ is centrally symmetric (this is always true when the filter $h$ is real-valued). Then $C$ is a symmetric real-valued matrix. On top of this we assume that $C$ is invertible, which requires that $|\hat h_i|>0$. Under these assumptions, the solution of (\ref{eq.optim}) may be obtained with the Lagrange's multipliers method, with
{\begin{equation}
\mathcal{L}(x,\lambda)=\frac{1}{2}x^{T}\cdot C\cdot x-\lambda^{*} \cdot(M\cdot x-y).\label{eq.Lagrange}
\end{equation}
Setting the derivatives of $\partial \mathcal{L}/\partial x_i$ to zero, together with Eq.~(\ref{eq.measurement}) give the following solution to the optimization problem~(\ref{eq.optim}):
\begin{equation}
x_0=P \cdot y,\label{eq.fast_reconstr}
\end{equation}
with 
\begin{equation}
P = C^{-1}\cdot M^{*}\cdot(M\cdot C^{-1}\cdot M^{*})^{-1},\label{eq.reconstr}
\end{equation}
where $C^{-1}=F^{*}\cdot \hat C^{-1}\cdot F$ and ${\hat C^{-1}}_{i,j}=\delta_{i,j} |\hat h_i|^{-2}$. The same result (\ref{eq.reconstr}) may be rewritten with the help of the Moore-Penrose matrix pseudoinverse $A^{+}=A^{*}\cdot(A\cdot A^{*})^{-1}$. By introducing a new diagonal matrix $\hat \Gamma=\hat C^{-1/2}$  we get
\begin{equation}
P =F^{*}\cdot\hat \Gamma\cdot F\cdot(M\cdot F^{*}\cdot\hat \Gamma \cdot F)^{+},\label{eq.pi_reconstr}
\end{equation}
where ${\hat \Gamma}_{i,j}=\delta_{i,j} |\hat h_i|^{-1}$. Either of the two formulas~(\ref{eq.reconstr}) or~(\ref{eq.pi_reconstr}) may be used equivalently to calculate matrix $P$. The pseudoinverse is usually calculated with the singular value decomposition which requires $k^2\cdot n$ multiplications. On the other hand, matrix inversion by Gaussian elimination requires only $k^3$ multiplications. However, for large matrices, the cost of~(\ref{eq.reconstr}) becomes dominated by matrix multiplication (with $k^2\cdot n$ multiplications needed to calculate $M\cdot M^{*}$) and is similar to that of~(\ref{eq.pi_reconstr}).
Still, we have used formula~(\ref{eq.reconstr}) since in practice it was almost twice faster than~(\ref{eq.pi_reconstr}) for the calculations presented in this paper. This result is obviously affected by many factors including the efficiency of parallel processing or memory bandwidth.  We have also neglected here the cost of multiplication by $F$ or $C^{-1}$, assuming the use of Fast Fourier Transform algorithm.
In any case, whichever of the formulas ~(\ref{eq.reconstr}) or~(\ref{eq.pi_reconstr}) is chosen, the calculation of $P$ is computationally demanding. Fortunately, after $P$ has been precalculated and stored in memory, $x_0$ may be reconstructed with Eq.~(\ref{eq.fast_reconstr}) in only $k\cdot n$ multiplications, which is interesting for real time image reconstruction. 

The proposed reconstruction algorithm is non-adaptive. It is also not based on the $\ell^1$ norm used in compressive sensing, so it does not tend to find a sparse representation of the image. Therefore, in a compressive measurement it is crucially important to use sampling which by itself is an image compression method, rather than to rely on the incoherence~\cite{Candes:invprob_23_969} between the sampling and compression bases. For instance, the sampling functions may be chosen to have a similar power spectrum to the typical power spectrum of real-world images with dominant low-frequency information. Then, optimizing the criterion (\ref{eq.crit}) may be seen as a way to regularize the result. Depending on the choice of the filter $h$, various features of the reconstructed image may be enhanced or suppressed, as long as this is not in contradiction with the measurement~(\ref{eq.measurement}). 

In this paper we consider a simple generalization of the criterion~(\ref{eq.crit}) by defining a similar composite criterion.  The modified criterion adds together the contributions from the norms obtained with different filters
\begin{equation}
E(x)= \sum_p {\alpha^{(p)} \| x * h^{(p)} \|^2},\label{eq.otf_crit}
\end{equation}
where the coefficients $\alpha^{(p)}>=0$  are responsible for weighting,  in a trade-off sense, the convex criteria obtained with every filter $h^{(p)}$. With the modified criterion, we have to adjust the definition of matrix $C$, which now becomes $C=\sum_p C^{(p)}$ with ${\hat C^{(p)}}_{i,j}=\delta_{i,j} |\hat h^{(p)}_i|^2$. This yields the definitions of $\hat C^{-1}$ and $\hat \Gamma$ necessary for Eqs.~(\ref{eq.reconstr}) and~(\ref{eq.pi_reconstr}),
\begin{equation}
\hat \Gamma_{i,j}=\frac{\delta_{i,j}}{\sqrt{ \sum_p\left| \alpha^{(p)}\cdot\hat h^{(p)}_i\right|^{2}} }
\textnormal{ and } \hat C^{-1}_{i,j}=\hat \Gamma_{i,j}^2, \label{eq.otf_Gamma}
\end{equation}

We will further assume that the composite criterion~(\ref{eq.otf_crit}) is constructed with the following filters: discrete gradient filters $h^{(1)}=\partial_x$, $h^{(2)}=\partial_y$, a penalty filter for high spatial frequencies $\hat h^{(3)}=\sqrt{\omega_x^2+\omega_y^2}$, and an identity filter $h^{(4)}=\delta(x,y)$. Gradient filters, whose transfer functions are $\hat h^{(1)}=\sin(\omega_x)$ and $\hat h^{(2)}=\sin(\omega_y)$, are responsible for minimizing image nonuniformity. The high frequency penalty filter puts a preference for a typical $1/|\omega|$ image spectrum. Finally, $h^{(4)}$ is needed to remove singularities from~(\ref{eq.otf_Gamma}). 
 \begin{figure}
\centering
\includegraphics[width=0.9\textwidth]{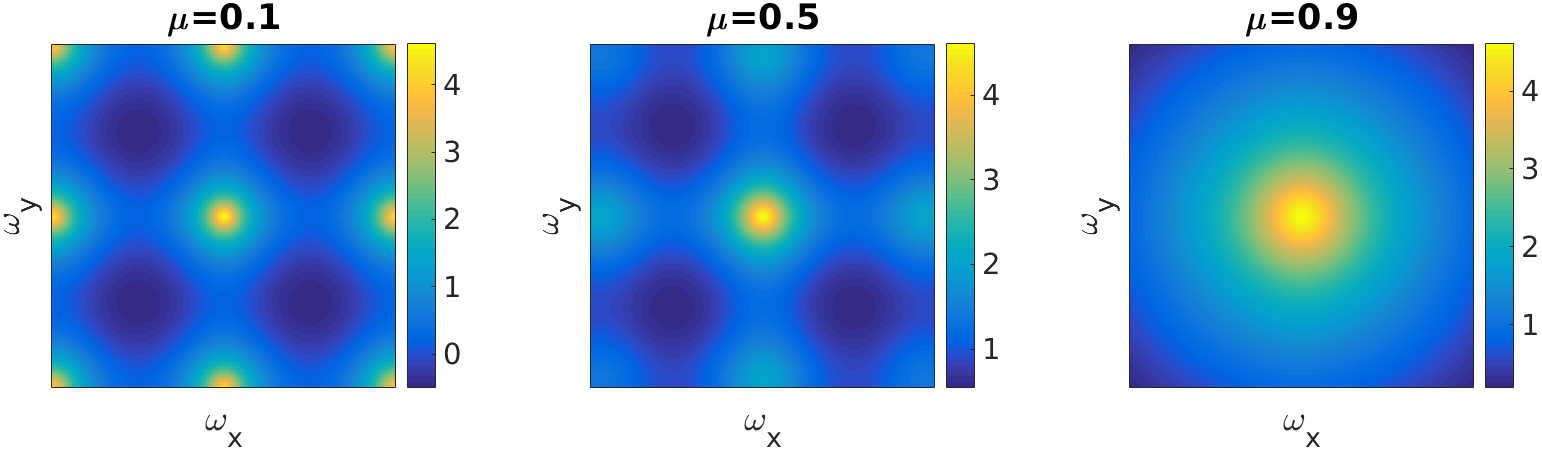}
\caption{Filter $\hat \Gamma(\omega_x,\omega_y)$ defined in Eq.~(\ref{eq.otf_Gamma_with_mu}) for various values of the parameter $\mu$ (logarithmic scale).\label{fig.gamma}}
\label{fig.filters}
\end{figure}
 
In the numerical part of the work we show the image recovery results as a function of a single parameter $\mu\in[0,1]$ (instead of $\alpha^{(p)}$). With our four filters $h^{(p)}$, $p=1,...,4$, we have the following form of $\hat \Gamma$,
\begin{equation}
\hat \Gamma_{i,j}=\frac{\delta_{i,j}}{\sqrt{ (1-\mu)^2 \left( \sin^2(\omega_x)+ \sin^2(\omega_y)\right) +\mu^2 \frac{\omega_x^2+\omega_y^2}{2\pi^2} +\varepsilon}},\label{eq.otf_Gamma_with_mu}
\end{equation}
where $(\omega_x,\omega_y)$ is the spatial frequency corresponding to index $i$, and $\omega_{x/y}\in(-\pi,+\pi)$. A small constant of  $10^{-5}$ is assigned to $\varepsilon$. Sample spectra $\hat \Gamma(\omega_x,\omega_y)$ are plotted in Fig.~\ref{fig.gamma} for various values of the parameter $\mu$.
We note that the proposed method converges to the pseudoinverse ($P\approx M^{+}$), when $\varepsilon>>1$.

To summarize this section, we reconstruct the image $x_0$ from the measurement $y$ using Eq.~(\ref{eq.fast_reconstr}), where the matrix $P$ is precalculated in advance with Eq.~(\ref{eq.reconstr}) or~(\ref{eq.pi_reconstr}), and $\hat \Gamma(\mu)$ is defined as (\ref{eq.otf_Gamma_with_mu}). Straightforward pseudoinverse-based reconstruction on the other hand is based on Eq.~(\ref{eq.fast_reconstr}) with the matrix $P=M^{+}$. 

\section{Sampling protocols} 
\label{seq:samp_protocols}
The FDRI reconstruction method presented in the previous section may be used with either orthogonal or nonorthogonal sampling.
 In this paper we consider three compressive sampling protocols, namely the discrete cosine transform (DCT) based sampling, Walsh-Hadamard sampling, and Morlet-wavelet based sampling, which all consist of small ($3\%$-$6\%$) subsets of the complete bases sets. 
We note that orthogonality is usually not preserved by binarization, which is needed for displaying the functions on DMD modulators. 
For instance, the DCT basis or Fourier basis binarized for use with DMD displays are no longer orthogonal. Sub-griding, multiple exposure, or sophisticated binarization algorithms~\cite{Zhang:scirep_7_12029} have been used to approximate the continuous functions with binary patterns more accurately. However, as we show, it may be sufficient to apply binarization directly to the continuous functions and to improve the reconstruction method instead. Other useful nonorthogonal bases interesting for single pixel detection are based on nonorthogonal wavelet bases or on speckle patterns. As for the orthogonal and inherently binary Walsh-Hadamard basis, the proposed reconstruction method may improve the reconstruction quality at a low compression ratio $k/n$, compared to the direct reconstruction with the inverse transform or a straightforward application of pseudoinverse ($x_0\approx M^{+}\cdot y$). 

\begin{figure}
\centering
\includegraphics[width=0.9\textwidth]{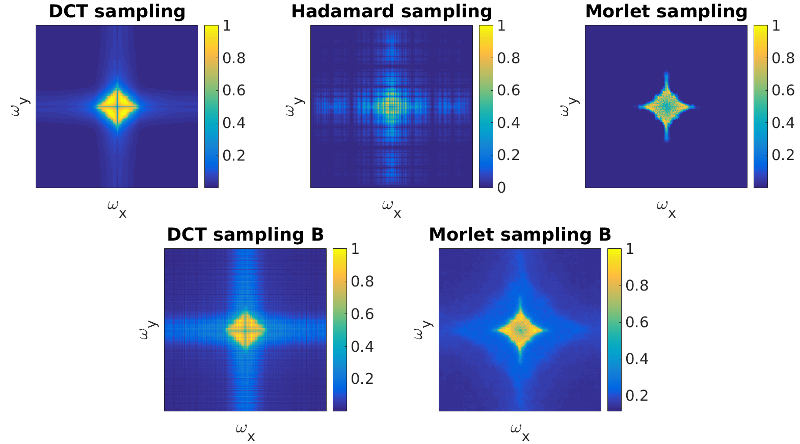}
\caption{Fourier spectra of sampling functions used in various sampling schemes ($\sum_i|\hat m_i|/\|\hat m_i\|$), where $m_i$ are the rows of the measurement matrix $M$, and the caret denotes the 2D Fourier transform. B denotes binarized sampling functions.}
\label{fig.sampling}
\end{figure}
A novel kind of nonorthogonal sampling for single pixel imaging has been recently proposed, with sampling functions obtained as a convolution between the real part of Morlet wavelets and realizations of white Gaussian noise~\cite{Czajkowski:scirep_8_466}. Therefore, let us now shortly recall its principles. 
The Morlet-based sampling proposed in~\cite{Czajkowski:scirep_8_466} is equivalent to a uniform random sampling of the representation of the image in the feature space constructed with Morlet wavelets. A feature space is built out of vectors, whose elements correspond to specific features of images. Simple features may be  associated with spatial and frequency contents of an image. Here, the feature space is constructed using Morlet wavelets and its part is preselected based on the analysis of an image database.
Morlet wavelets, also known as Gabor wavelets, consist of products of Gaussian functions with exponential functions. In two dimensions, a Morlet wavelet is equal to
\begin{equation}
g_{\sigma,n_p,\theta}(x,y)=Ne^{-\frac{x^2+y^2}{2\sigma^2}}(e^{i(\pi n_p/2\sigma)(x \cos(\theta)+y \sin(\theta))}-\kappa),
\label{eq.morlet}
\end{equation}
where the constants $\kappa$ and $N$ assure that the wavelet function $g$ is normalized $\|g\|=1$ and has zero mean $\overline{g}=0$. Parameters $\sigma,n_p,\theta$ are related to the size of the Gaussian envelope, number of periods within the envelope, and the orientation of modulation. The Morlet wavelet based sampling functions considered in this paper are obtained as convolutions of (\ref{eq.morlet}) with realizations of white Gaussian noise. 
Similarly as with the DCT basis, the binarized Morlet-based sampling functions considered here are obtained from the continuous ones with a direct thresholding at the mean value of the continuous function.

 Morlet wavelets have well established unique Fourier domain properties that decide upon their suitability for applications which require joint image description in the image and Fourier spaces. In fact, visual information is often analyzed in either of or in both of these two spaces. The Fourier transform of a Morlet wavelet is also a Morlet wavelet. Moreover, two Morlet wavelets constituting a Fourier function pair equalize the uncertainty  principle~\cite{Daugman:josaa_2_1160}. Therefore, Morlet wavelets are a preferable choice for creating an image representation maintaining the best possible resolution trade-off between image and Fourier spaces.  
 In spite of the importance of Morlet wavelets in quantum mechanics with similarity to quantum coherent states, their role in human and artificial vision, and usefulness for multi-scale analysis shared with other wavelet representations, they have also drawbacks which limit their applications. In particular, they are in general not orthogonal, the expansion into Morlet wavelets is not unique and is computationally more costly than finding the Fourier representation, the kernel of which in fact consists of a particular orthogonal subset of Morlet wavelets.
 
 We have selected the sampling functions proportionally to the frequency of their occurrence in a compressive decomposition of an image database. 
For the first two orthonormal functions, a set of $49$ images has been decomposed either in DCT or in Walsh-Hadamard basis and the functions with maximal average contribution to the respective transform have been selected.
Since Morlet wavelets do not form an orthogonal basis,
the same approach cannot be used. In this case, we find the optimal range of values for $\frac{n_p}{\sigma}$ ratio (see Eq.~(\ref{eq.morlet})) by decomposing a set of $49$ images in Discrete Fourier Transform (DFT) basis. Then, we calculate $n_p$ and $\sigma$ using the following equation:
\begin{equation}
 \sigma=\sigma_{min}+(\sigma_{max}-\sigma_{min}) \frac{\omega_{max}-\omega}{\omega_{max}-\omega_{min}},
\end{equation}
where $\omega=\frac{n_p}{2\sigma}$. This relation results in the maximal $\sigma$ being obtained for minimal spatial frequency $\omega$ and minimal sigma being obtained for maximal spatial frequency leading to an efficient sampling protocol.
 The averaged spatial spectra of the sampling functions obtained this way are compared in Fig.~\ref{fig.sampling}. These spectra are dominated by low-frequency contents, and binarization clearly extends them towards higher spatial harmonics. We note that without weighting of the sampling functions, these spectra do not need to match an average spectrum of the images, and also clearly they differ between each other.
 
 \section{Numerical results}
 
 \begin{figure}
\centering
\includegraphics[width=\textwidth]{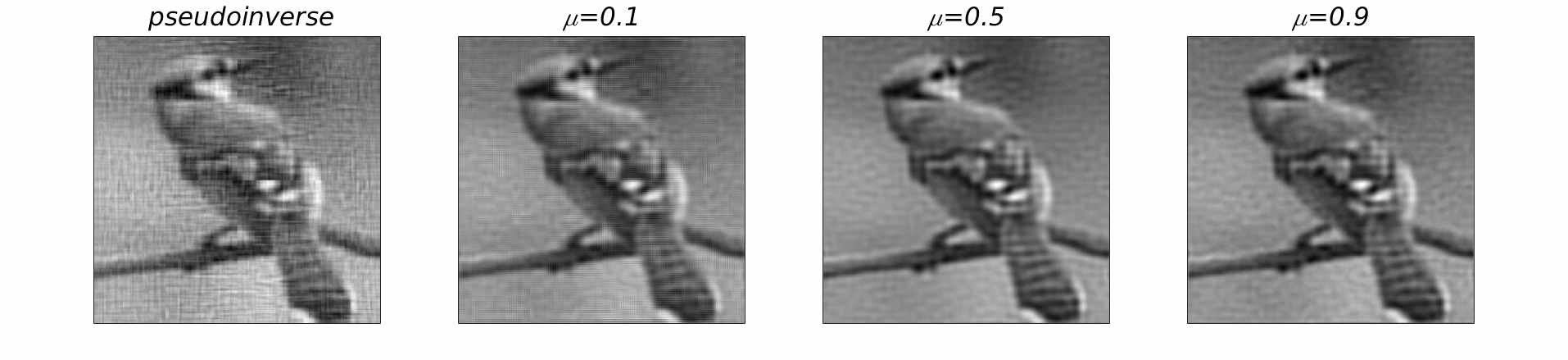}
\caption{Numerical simulations of compressive imaging using $3\%$ binarized Discrete Cosine Transform (DCT) basis functions. The results of the Fourier domain regularized inverse algorithm with varying parameter $\mu$ are compared with a  reconstruction obtained using pseudoinverse of the measurement matrix.}
\label{fig.num.mu}
\end{figure}

Numerical simulations of computational imaging using various sampling protocols have been conducted to evaluate the performance of the proposed FDRI reconstruction algorithm. 
In Fig.~\ref{fig.num.mu} we compare reconstructions obtained with different values of parameter $\mu$ (see Eq.~(\ref{eq.otf_Gamma_with_mu}) ) together with a reconstruction based on the pseudoinverse of the measurement matrix. For $\mu$=0, the filter corresponds to discrete gradient norm, while if it is close to unity it corresponds to penalizing high spatial frequencies. The presented test image was sampled with binary DCT functions at $3\%$ compression ratio. In case of  the pseudoinverse-based reconstruction, high level of undersampling and binarization of the sampling functions result in undesired artifacts and loss of the image quality. With the FDRI method the quality of the reconstructed image is visibly improved for all considered values of parameter $\mu$. In the later part of this work, we choose $\mu=0.5$, which balances filters contributing to the composite criterion in Eq.~(\ref{eq.otf_crit}).

\begin{figure}
\centering
\includegraphics[width=\textwidth]{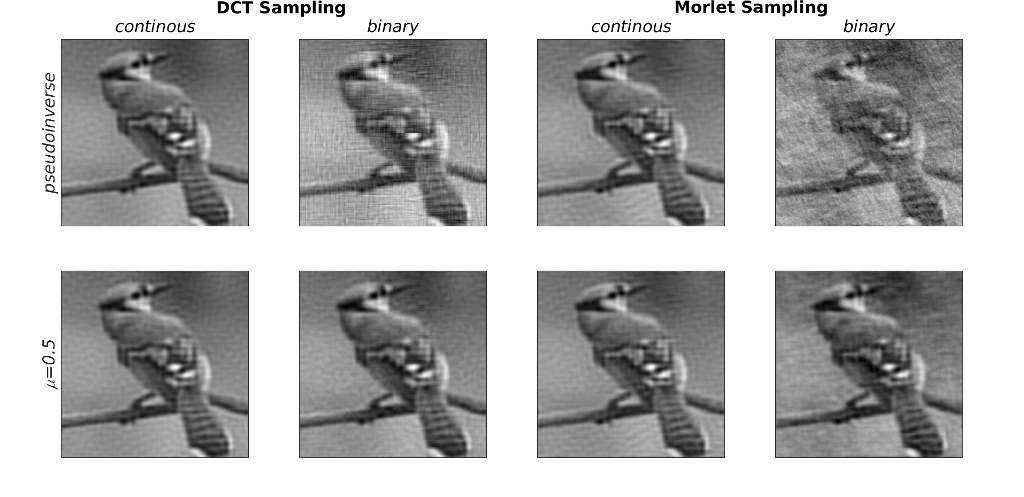}
\caption{Numerical simulation of compressive imaging presenting the influence of Fourier domain regularized inverse (FDRI) with $\mu=0.5$ for various sampling schemes at $3\%$ compression ratio. Compared to the direct pseudoinverse method (top row), the proposed reconstruction algorithm (bottom row) offers by far lower degradation of the image quality resulting from utilizing binarized sampling patterns. At the same time, when continuous sampling functions are used, both reconstruction methods lead to comparable image quality.}
\label{fig.num.sampling}
\end{figure}

In the second simulation, we analyze the performance of FDRI method, when either continuous or binary sampling protocols are used.
For continuous sampling functions composed of either Morlet wavelets convolved with white noise or DCT basis functions, reconstructions of comparably good quality are obtained for both pseudoinverse and FDRI recovery method (see Fig.~\ref{fig.num.sampling}). 
However, binarization of the sampling function leads to a major deterioration of the  images recovered with pseudoinverse, which is caused by drastic modifications in both spatial and spectral properties of the sampling patterns. On the other hand,  the quality of images reconstructed with FDRI method is similar for both cases of sampling with continuous and binarized functions. The FDRI method tailors the Fourier spectrum of the solutions and therefore may diminish the error introduced by using binary, instead of grayscale, sampling patterns.

\begin{figure}
\centering
\includegraphics[width=0.9\textwidth]{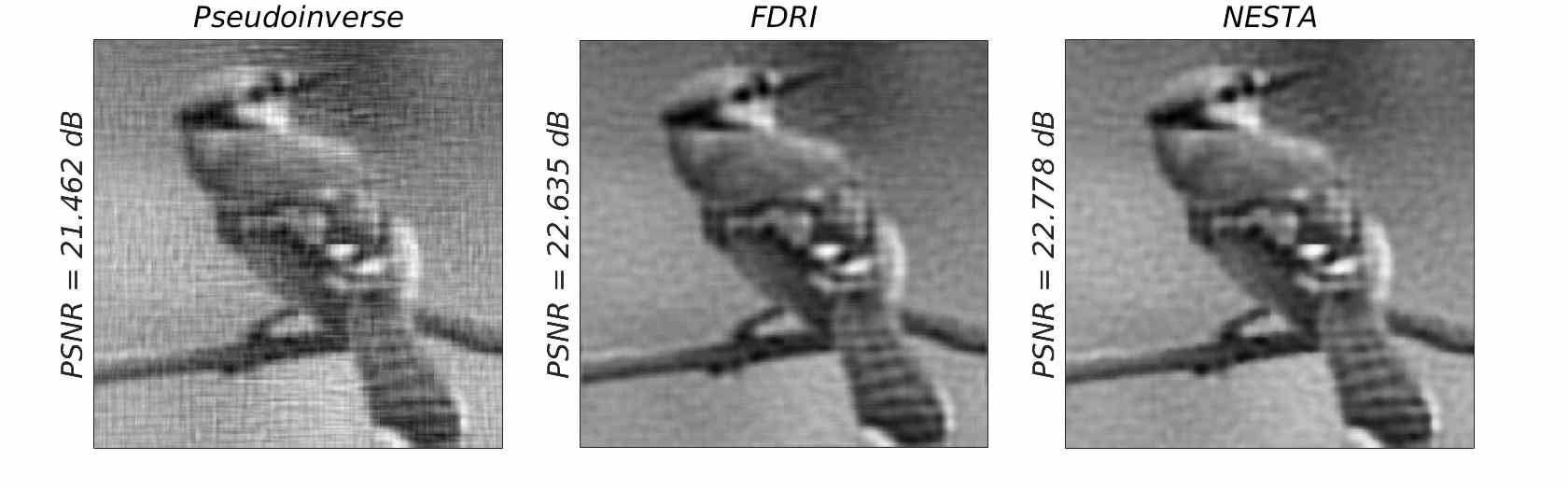}
\caption{Numerical simulations of compressive imaging presenting a comparison between various reconstruction methods for Discrete Cosine Transform sampling functions at $3\%$ compression ratio. The peak signal-to-noise ratio (PSNR) values indicate that the proposed method (FDRI) enables reaching similar performance to that of total variation  regularization (NESTA), while pseudoinverse based reconstruction is discrepant from the other two.}
\label{fig.num.recon}
\end{figure}

\begin{figure}
\centering
\includegraphics[width=\textwidth]{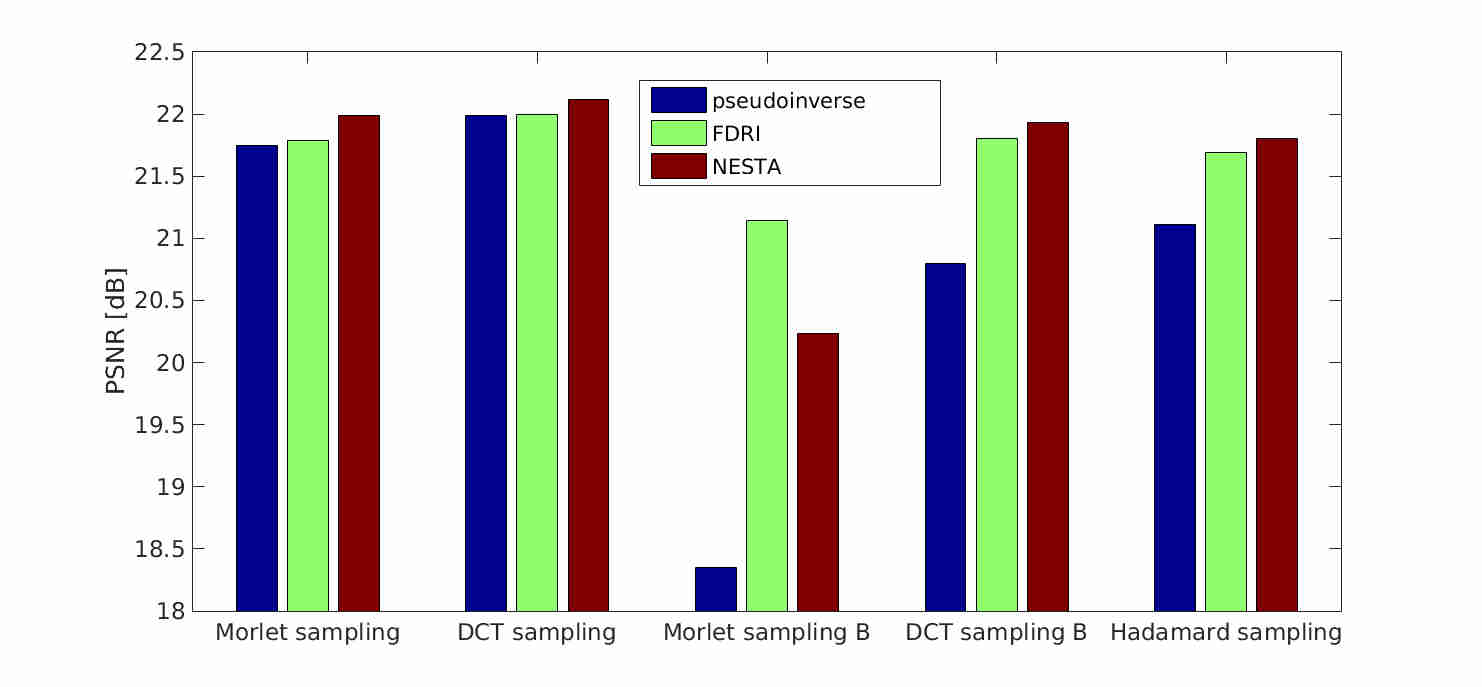}
\caption{Comparison of peak signal-to-noise ratio (averaged over $49$ test images) obtained for various sampling and reconstruction methods. The improvement gained by using the FDRI reconstruction method is especially visible for the binary sampling methods denoted with B.}
\label{fig.num.barplot}
\end{figure}

Finally, we compare the image reconstructions obtained with the FDRI method  with NESTA\cite{Becker:sciam2009} package, which utilizes total variation (TV) minimization to reconstruct the image. TV optimization is one of the most common reconstruction methods used in single-pixel imaging. 
We show in Fig.~\ref{fig.num.recon}, that the FDRI method allows to reconstruct compressively measured images with quality similar to the one offered by NESTA. In the presented example, for sampling with $3$\% binary DCT patterns, the peak signal-to-noise ratio (PSNR) for these two methods differs by only $0.2$~dB in favor of NESTA.
 At the same time, a considerable advantage offered by the FDRI method is reconstruction time. Since only single matrix by vector multiplication is required, the image reconstruction is fast. For the image resolution of $256\times 256$ and 3\% compression ratio, it takes approx. $0.09$~s on a medium-class PC with the use of single precision arithmetics. 
 Reconstruction conducted with NESTA in the same conditions requires $7$~s of computations, which precludes the methods based on TV minimization  from applications in real time single-pixel video imaging. 

As a summary, a set of $49$ test images have been reconstructed with all three methods, namely: FDRI, pseudoinverse and NESTA and analyzed in terms of PSNR for several different sampling protocols (see Fig.~\ref{fig.num.barplot}).
In most cases, the use of NESTA results in only marginally higher PSNR than it is obtained with FDRI method. When binarized sampling patterns are used, FDRI offers considerable improvement in reconstruction quality in comparison with  pseudoinverse. For binary Morlet sampling protocol, the FDRI method outperforms also the recoveries obtained with TV minimisation.  On the other hand, for continuous sampling protocols, PSNR values obtained with pseudoinverse and FDRI methods are similar, while a minor increase still can be acquired with NESTA. The highest qualities of the reconstructed images are obtained for continuous sampling methods. However, when only binary sampling functions are considered, the use of binarized DCT functions together with the FDRI or NESTA reconstruction method seems to offer the highest PSNR values. Only slightly worse results are obtained with Hadamard sampling.

\section{Experimental results}
\begin{figure}
\centering
\includegraphics[width=0.65\textwidth]{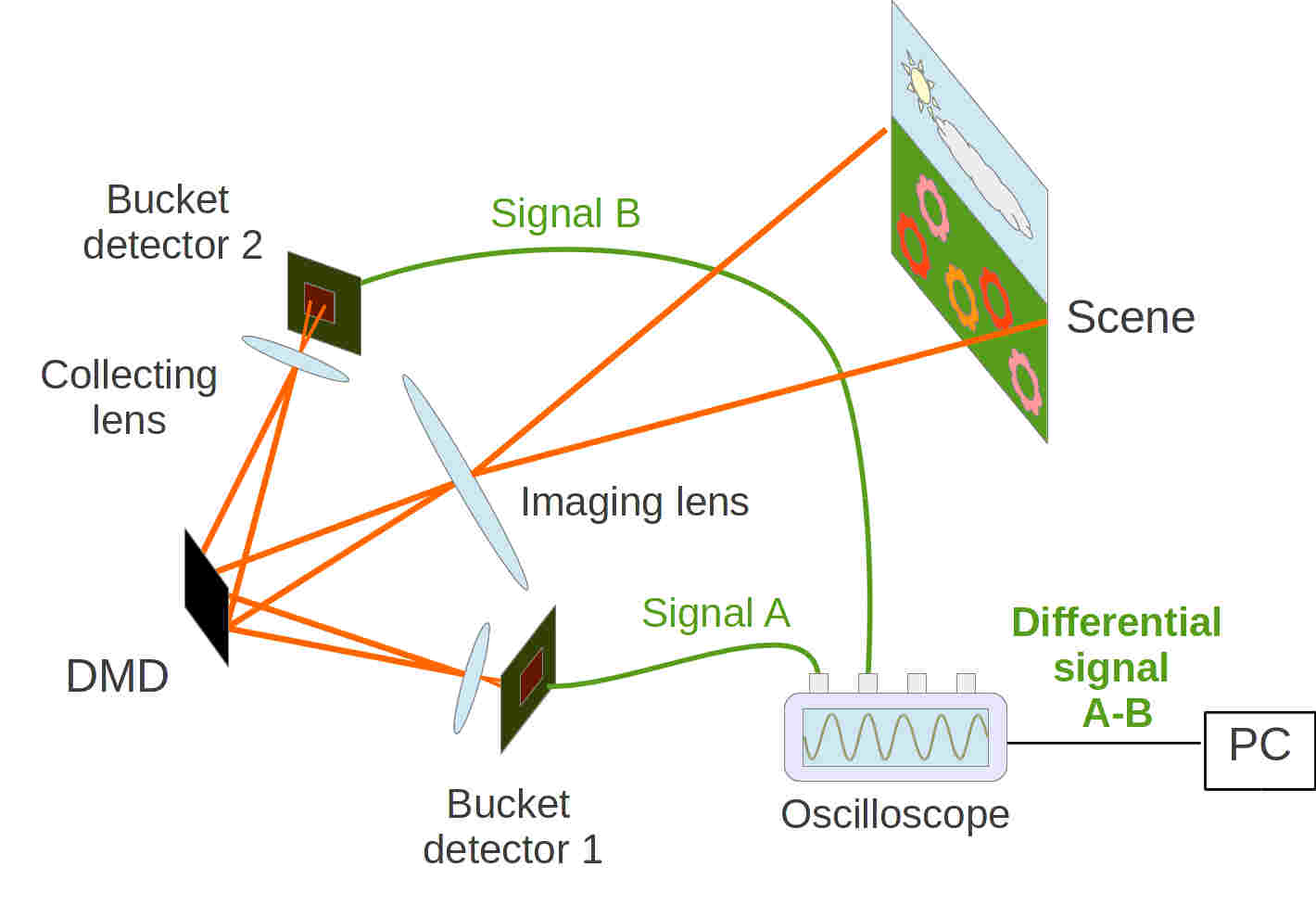}
\caption{Schematic view of the single-pixel camera with DMD spatial modulator used for image sampling. Differential photo-detection  with two bucket detectors gathering simultaneously the complementary optical signals reflected from DMD is used to increase experimental SNR.}
\label{fig_setup}
\end{figure}
\begin{figure}
\centering
\includegraphics[width=0.9\textwidth]{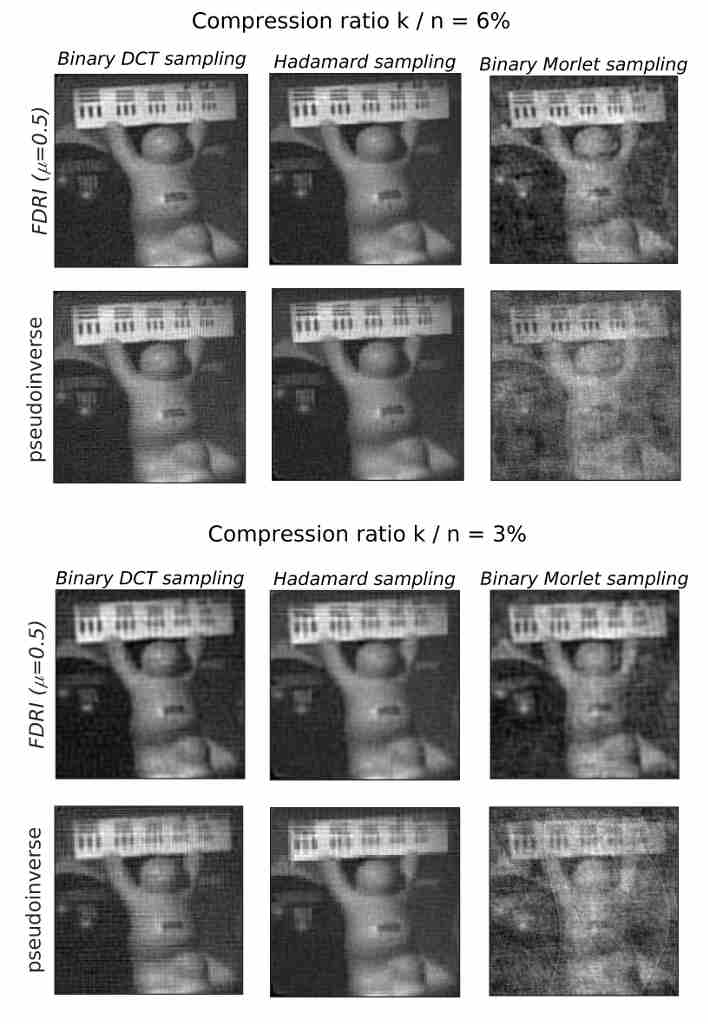}
\caption{Comparison of reconstructions obtained with FDRI and pseudoinverse method for experimental data measured with an optical single-pixel camera with different types of sampling protocols at resolution of $256 \times 256$ and compression ratios of $6\%$ (top 6 images) and $3\%$ (bottom 6 images). }
\label{fig_exp_result}
\end{figure}

We have verified our results experimentally with an optical single-pixel camera setup, schematically illustrated in Fig.~\ref{fig_setup}. The projection of a three-dimensional scene is sampled with binary patterns displayed by a DMD with $1024 \times 768$ pixel resolution and maximum sampling rate of $22$~kHz. The signal is then measured using a differential photo-detection technique~\cite{Telescopic_scirep2014Yu,Balanced_scirep2016Lancis}, with two photodiodes simultaneously collecting light reflected from the areas of the DMD set to both \textit{on} and \textit{off} states. 
This approach, as compared to typical single-pixel cameras equipped with only one photodetector, significantly increases the experimental signal-to-noise ratio by compensating effects of background illumination, variations of light intensity over time and signal oscillations induced by electronic equipment. 
The signals are sampled with a $15$-bit digital oscilloscope at the rate of $7.8$~MS/s and the measurements are streamed on-the-fly to a PC, where the reconstruction of the video is conducted in parallel. 
Together with a cost-efficient image recovery method, in which only a single product of a precalculated matrix with a vector is needed per image reconstruction, the camera is capable of continuous sampling and reconstructing video images in real time, with the frame rate dependent only on the number of sampling patterns used in a compressive measurement. For instance, with resolution of $256 \times 256$ and $3\%$ compression ratio, the video frame rate exceeds $11.3$~Hz.

\begin{figure}
\centering
\includegraphics[width=0.8\textwidth]{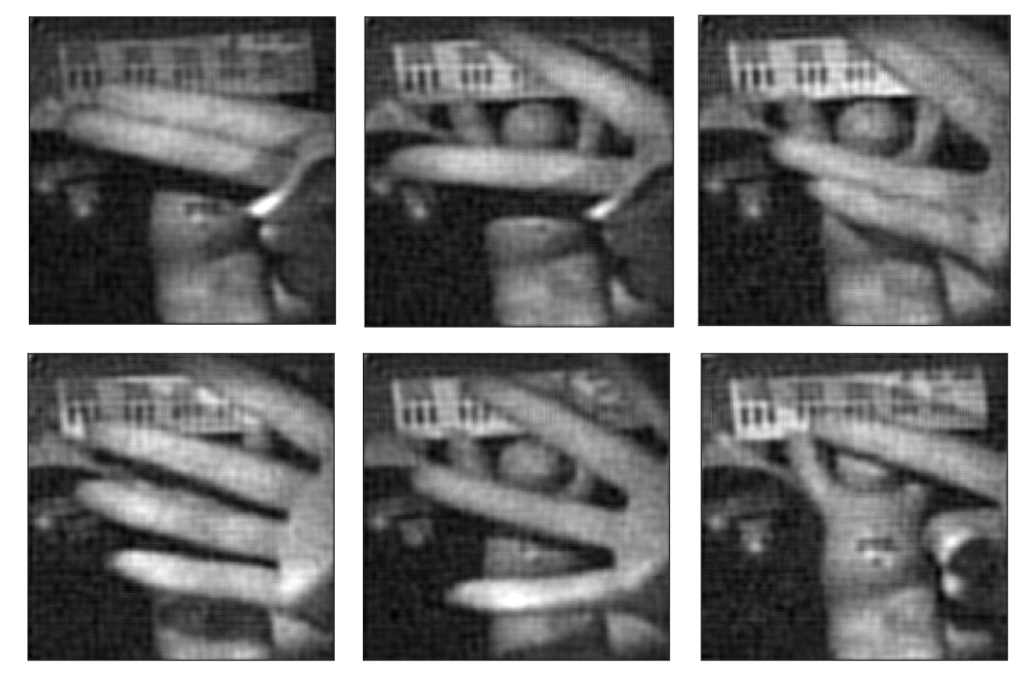}
\caption{Several exemplary video frames recorded with our optical single-pixel camera at the rate of $11.3$~Hz and reconstructed on-the-fly using the proposed FDRI algorithm. The sampling patterns consist of $3\%$ binarized discrete cosine transform (DCT) basis functions with resolution $256 \times 256$. For full video see Visualization 1.}
\label{fig_exp_film}
\end{figure}

The sampling functions used in the experiment were chosen in the same manner as discussed in Section~\ref{seq:samp_protocols}, however  only binary functions have been considered.
Although the effect of grayscale patterns may be also achieved with DMDs by time-division multiplexing, only binary patterns are displayed at the full speed, which plays a crucial role in real-time imaging. 
The resolution of the sampling functions was set to $256 \times 256$, while the patterns were resized to fill the whole area of the DMD display. In each set of sampling functions a single white pattern (with all pixels equal $1$) was included for the purpose of synchronization between the DMD and the data acquisition unit. 
In Fig.~\ref{fig_exp_result} we present the reconstructions of a single representative video frame obtained with our camera using three different types of sampling functions at compression ratios of $3\%$ and $6\%$.
The reconstructions calculated using the proposed FDRI method are compared with recoveries based on the pseudoinverse of the measurement matrix. The improvement of the reconstruction quality with FDRI method is especially robust for highly compressive measurements, when the filtering properties of this approach allow for smoothing the undesired artifacts resulting from the lossy compression accompanying the measurement. At the same time, the sharpness of edges and information about small details in the image, such as the resolution chart, are preserved with high accuracy.
Finally, in Fig.~\ref{fig_exp_film} we illustrate a selection of video frames recorded by our single-pixel camera using $3\%$ of DCT basis functions for sampling the image at the frame rate of $11.3$~Hz. Video frames have been reconstructed on the fly using the FDRI algorithm with parameter $\mu = 0.5$. Full video has been included in Visualization 1.

\section{Conclusions}
We have proposed a fast, linear, closed-form reconstruction method for single pixel real-time video imaging and we have demonstrated experimentally its feasibility to reconstruct good quality video with $256\times 256$ resolution and $11$~Hz frame rate on the fly.
The reconstruction method is based on the generalized inverse of the rectangular 
measurement matrix. It is obtained by constrained optimization of a quadratic criterion, which is designed to select solutions with desirable features. 
The criterion is defined using a combination of spatial filters corresponding  to minimization of discrete gradients of an image and a penalty filter for high spatial frequencies. The proposed method shares 
a major advantage with the direct pseudoinverse-based reconstruction. Namely, the recovery of an image requires only a single-step calculation with linear numerical cost as a function of the number of measurements. Therefore, the reconstruction is extremely fast and may be conducted on the fly 
with a single-pixel video camera equipped with a state-of-the-art DMD. For instance, in our experiment the image has been sampled by a DMD at the rate of $22$~kHz with binary patterns obtained from $3\%$ of the discrete cosine transform basis. On the other hand, the reconstruction matrix needs to be calculated beforehand and stored in memory, which may be challenging for imaging at high resolution. Fortunately, it is possible to reduce the memory cost by choosing a well-designed sampling protocol, which allows for highly compressive measurement.  In our case, this is achieved by selecting the sampling functions proportionally to the frequency of their occurrence in a compressive decomposition of an image database.

The criterion included in the definition of the reconstruction procedure provides a better regularization of the inversion of the measurement matrix than the direct pseudoinverse. Therefore, it leads to much higher quality of the reconstructed images, with reduced amount of noise and numerical artifacts.
We prove that the proposed reconstruction method is comparable with solvers based on the optimization of the total variation, such as NESTA, in terms of the PSNR of the recovered images, while the reconstruction time in the same conditions is reduced multiple times with our method. 
Moreover, the proposed reconstruction method remains valid for nonorthogonal sampling functions.
Nonorthogonal functions are obtained for instance by binarizing continuous functions, such as DCT or Fourier basis or Morlet wavelets convolved with white noise. While binary sampling offers better compatibility with DMDs, it usually has different spectral properties from the original continuous functions and therefore leads to some degradation of the quality of images reconstructed with pseudoinverse. We have shown that the proposed method is well suited to mitigate this effect.

\section*{Acknowledgments}
The authors acknowledge partial financial support from the Narodowe Centrum Nauki (grant No. UMO-2014/15/B/ST7/03107).




\end{document}